\documentclass[%
reprint,
superscriptaddress,
amsmath,amssymb,
aps,longbibliography,
prl,
twocolumn,
floatfix,
]{revtex4-2}

\usepackage{xr}

\makeatletter
\newcommand*{\addFileDependency}[1]{%
\typeout{(#1)}
\@addtofilelist{#1}
\IfFileExists{#1}{}{\typeout{No file #1.}}
}\makeatother

\newif\ifshowrev
\showrevfalse %

\ifshowrev
  \newcommand{\rev}[1]{{\color{red}#1}}
\else
  \newcommand{\rev}[1]{#1}
\fi

\usepackage{graphicx}%
\usepackage{verbatim}
\usepackage[version=4]{mhchem}
\usepackage{xcolor}
\usepackage{physics}
\usepackage[normalem]{ulem}
\usepackage{pdfpages}
\usepackage{pgffor}
\usepackage{soul}
\usepackage{booktabs}
\usepackage{multirow}
\usepackage{caption}
\usepackage{subcaption}
\usepackage{xcolor}

\makeatletter
\AtBeginDocument{\let\LS@rot\@undefined}
\makeatother

\begin{document}

\title{Learning the action for long-time-step simulations of molecular dynamics}

\author{Filippo Bigi}
\email{filippo.bigi@epfl.ch}
\affiliation{Laboratory of Computational Science and Modeling, Institut des Mat\'eriaux, \'Ecole Polytechnique F\'ed\'erale de Lausanne, 1015 Lausanne, Switzerland}

\author{Johannes Spies}
\affiliation{Laboratory of Computational Science and Modeling, Institut des Mat\'eriaux, \'Ecole Polytechnique F\'ed\'erale de Lausanne, 1015 Lausanne, Switzerland}

\author{Michele Ceriotti}
\email{michele.ceriotti@epfl.ch}
\affiliation{Laboratory of Computational Science and Modeling, Institut des Mat\'eriaux, \'Ecole Polytechnique F\'ed\'erale de Lausanne, 1015 Lausanne, Switzerland}

\newcommand{\FB}[1]{{\color{teal} #1}}
\newcommand{\FBcancel}[1]{{\color{teal} \sout{#1}}}
\newcommand{\MC}[1]{{\color{blue} #1}}
\newcommand{\MCcancel}[1]{{\color{blue} \sout{#1}}}

\date{\today}

\begin{abstract}
The equations of classical mechanics can be used to model the time evolution of countless physical systems, from the astrophysical to the atomic scale.
Accurate numerical integration requires small time steps, which limits the computational efficiency -- especially in cases such as molecular dynamics that span wildly different time scales.
Using machine-learning (ML) algorithms to predict trajectories allows one to greatly extend the integration time step, at the cost of introducing artifacts such as lack of energy conservation and loss of equipartition between different degrees of freedom  of a system. 
\rev{We propose learning data-driven structure-preserving (symplectic and time-reversible) maps to generate long time-step classical dynamics and show that this method is equivalent to learning the mechanical action of the system of interest. These models can be learned based on short reference trajectories, and be transferred across thermodynamic conditions and chemical composition.}
We show that an action-derived ML integrator eliminates the pathological behavior of non-structure-preserving ML predictors, and that the method can be applied iteratively, serving as a correction to computationally cheaper direct predictors.

\end{abstract}

\maketitle

\newcommand{\bp}{\boldsymbol{p}}
\newcommand{\bq}{\boldsymbol{q}}
\newcommand{\bx}{\boldsymbol{x}}

\newcommand{\bdp}{\Delta\boldsymbol{p}}
\newcommand{\bdq}{\Delta\boldsymbol{q}}
\newcommand{\bdx}{\Delta\boldsymbol{x}}

\newcommand{\bpbar}{\Bar{\boldsymbol{p}}}
\newcommand{\bqbar}{\Bar{\boldsymbol{q}}}
\newcommand{\bxbar}{\Bar{\boldsymbol{x}}}

Simulating classical mechanical systems with high accuracy and efficiency is a long-standing challenge in computational physics~\cite{landau2024computational, leimkuhler2004simulating}. 
Traditional numerical methods typically rely on small time steps to propagate the equations of motion for the dynamical system in order to provide accurate integration. 
These small steps limit computational speed and scalability, especially for problems such as atomistic simulations, which feature a large gap between the time scale of the fastest motion and that of the slow collective transitions that determine relevant physical processes~\cite{klep+09cosb}.

Recent advances in machine learning offer promising alternatives by enabling the data-driven approximation of complex physical processes~\cite{brunton2020machine,degrave2022magnetic,reichstein2019deep,carleo2019machine,greydanus2019hamiltonian,rath2021symplectic,chen2021data,offen2022symplectic,jin2020sympnets}. 
For example, recent work on the machine-learning-driven prediction of molecular dynamics trajectories using long time steps has demonstrated the potential for a groundbreaking speedup of atomistic simulation workflows~\cite{mdnet, vlachas2021accelerated, trajcast, draldirectmd, bigi2025flashmd, dayhoff2025mlmd}. 
However, these methods do not preserve the geometric structure of the underlying Hamiltonian flow, leading to violations of conservation of energy, equipartition, and other fundamental physical laws, which hamper their use for rigorous scientific applications. 
Here, we investigate a parametrization approach to directly learn structure-preserving maps that approximate the long-time evolution of classical dynamical systems, with the goal of significantly increasing simulation time steps while maintaining geometric and physical fidelity.

The general class of problems we are interested in are those whose time evolution obeys Hamilton's equations
\begin{equation}\label{eq:hamilton}
    \frac{d \bp}{d t} = -\frac{\partial H}{\partial \bq}, \quad \frac{d \bq}{d t} = \frac{\partial H}{\partial \bp},
\end{equation}
where $\bq$ and $\bp$ indicate the position and momentum vectors, each having a dimensionality equal to the number of degrees of freedom in the mechanical system. Here and in the rest of this work, we will assume the Hamiltonian to be independent of time, which is appropriate for a closed system. For example, in many scientifically relevant problems, $H$ takes the form 
\begin{equation}\label{eq:hamiltonian}
    H(\bp, \bq) = \sum_{i=1}^F \frac{p_i^2}{2m_i} + V(\bq),
\end{equation}
where $m_i$ are the masses associated with each of the $F$ degrees of freedom and $V(\bq)$ is the potential energy of the system. This includes most classical systems from astronomy to molecular dynamics.

Given the ubiquity of this class of problems %
in many areas of mathematics and physics, many integration approaches have been developed for the numerical solution of Hamiltonian dynamics. 
Among these, algorithms which preserve specific geometrical properties of the exact Hamiltonian flow have been shown to possess desirable long-time behavior~\cite{hairer2010geometric}.
Symplectic integrators, in particular, preserve exactly, for any time step, a geometric term corresponding to an area element in $(\bp,\bq)$ space
(see the SI for a concise summary of textbook results in this field). 
Symplecticity guarantees the existence of a modified (or \emph{shadow}) Hamiltonian whose exact flow corresponds to the numerical solution to a very good approximation over very long times~\cite{benettin1994hamiltonian,hairer2010geometric}. 
This ensures that the numerically propagated system is also Hamiltonian and, since the modified Hamiltonian is close to the true one, long-time near-conservation of energy. 
Furthermore, time-reversible methods have the advantage that their modified Hamiltonian is equal to the real one up to second order in the time step, further improving their accuracy and energy conservation.

An alternative approach to generate the evolution of a Hamiltonian system is to consider it purely as a learning problem: given the momentum and position at time $t$, $\bp := \bp(t)$ and $\bq := \bq(t)$, one aims to predict the evolved values $\bp' := \bp(t+h), \bq' := \bq(t+h)$, where $h$ is a (potentially large) time step. 
As \rev{ mentioned earlier}, several works have recently shown machine-learning (ML) models that predict $(\bp',\bq')$ with high accuracy up to time steps that are two orders of magnitude longer than the stability limit of conventional integrators~\cite{trajcast,draldirectmd,bigi2025flashmd}, and that can be applied across large portions of chemical space for molecular and materials simulations~\cite{draldirectmd,bigi2025flashmd}.
Despite their accuracy, the resulting long-time trajectories are unstable because the model does not conserve energy. 
As discussed in Ref.~\cite{bigi2025flashmd}, this can be mitigated by rescaling and thermostatting the particle velocities, but the lack of an underlying Hamiltonian structure leads to the appearance of other artifacts, such as loss of equipartition, that are hard to monitor and correct.

We can take a different approach to the learning problem, using a \rev{parametrization} that preserves the structure of the Hamiltonian problem.
It is known that, under mild assumptions~\cite{hairer2010geometric}, any symplectic map $(\bp, \bq) \rightarrow (\bp', \bq')$ can be defined by a scalar generating function $S$, and vice versa. The generating function can be parametrized in a number of ways, the most common ones being
\begin{equation}
    S(\bq, \bq'), \,\, S^1(\bp', \bq), \,\, S^2(\bp, \bq'), \,\, S^3(\bpbar, \bqbar),
\end{equation}
where we have followed the notation in Ref.~\citep{hairer2010geometric}, and where $\bpbar = (\bp + \bp')/2, \, \bqbar = (\bq + \bq')/2$. Among these, we select the $S^3$ parametrization, because it is symmetric, it leads to a simple and elegant condition for time-reversible maps, and because the evaluation of the associated symplectic transformation is equivalent to the well-known implicit midpoint rule~\cite{hairer2010geometric}, whose practical implementation is discussed below.
More details on the choice of $S^3$ are available in the SI. A generating function in the form $S^3(\bpbar, \bqbar)$ defines the symplectic map as
\begin{equation}\label{eq:s3-predictions}
    \bdp = -\frac{\partial S^3}{\partial \bqbar}, \quad \bdq = \frac{\partial S^3}{\partial \bpbar},
\end{equation}
where $\bdp = \bp' - \bp$ and $\bdq = \bq' - \bq$.
Then, expressing $S^3$ as a neural network $\Tilde{S}^3$ leads to a generic parametrization of the symplectic map $(\bp, \bq) \rightarrow (\bp', \bq')$, and the neural network can be trained on $((\bp, \bq), (\bp', \bq'))$ pairs generated by a conventional small-time-step integrator. This \rev{formulation} ensures that the predicted time-evolution is symplectic, but not necessarily time reversible. To enforce this additional symmetry, which can be expressed as the constraint that $(-\bp', \bq') \rightarrow (-\bp, \bq)$, one needs to ensure that $S^3(\bpbar, \bqbar) = S^3(-\bpbar, \bqbar)$ (see SI).
This can be enforced without loss of generality if $S^3$ is represented by a neural network $\Tilde{S}^3$. Indeed, it is sufficient to symmetrize the neural network with respect to $\bpbar$, for example
\begin{equation}\label{eq:time-rev}
    \Tilde{S}^3(\bpbar, \bqbar) \leftarrow \frac{\Tilde{S}^3(\bpbar, \bqbar) + \Tilde{S}^3(-\bpbar, \bqbar)}{2}.
\end{equation}

Even though we will not consider the machine learning of variable step sizes in this work, it is instructive to discuss the dependence of the generating functions $S$ and $S^3$ on the time step $h$. Indeed, it can be shown~\cite{hairer2010geometric} that the time-dependent generating function $S(\bq, \bq', h)$ must satisfy the Hamilton-Jacobi partial differential equation:
\begin{equation}\label{eq:hamilton-jacobi}
    - \frac{\partial S}{\partial h} = H \Big(\frac{\partial S}{\partial \bq'}, \bq' \Big),
\end{equation}
and that it therefore corresponds to the action of the system.
Since $S^3$ is related to $S$ (up to a constant) by
\begin{equation}
    S^3(\bpbar, \bqbar, h) = \bpbar(h) \cdot \bdq(h) - S(\bq, \bq', h),
\end{equation}
learning the symplectic map $S^3$ generated by Hamiltonian flow effectively amounts to learning the action $S$ of the system. Here, the ``action'' $S(\bq, \bq', h)$ refers to Hamilton’s principal function, also known as Hamilton–Jacobi action, which corresponds to the minimized values of the action functional over paths starting at $\bq$, ending at $\bq'$, and taking time $h$ to do so.

The formal time-dependence of the neural network approximation $\Tilde{S}^3$ can also be used to rigorously establish the existence of a modified Hamiltonian for the simulations generated by it. 
Assuming that the training procedure of the neural network $\Tilde{S}^3(\bpbar, \bqbar, h)$ is a smooth and infinitely differentiable function of $h$ through the $h$-dependence of the training samples $((\bp, \bq), (\bp'(h), \bq'(h)))$, then we can consider $\Tilde{S}^3(\bpbar, \bqbar, h)$ itself as a smooth and infinitely differentiable function of $h$. 
Within this assumption (see Chapter 9 of Ref.~\citep{hairer2010geometric}), the discretized simulation follows the dynamics generated by a modified Hamiltonian, which confers it the favorable properties of Hamiltonian dynamics, including long-time conservation of energy, equipartition, etc.
\begin{figure}[b]
    \centering
    \includegraphics[width=\linewidth]{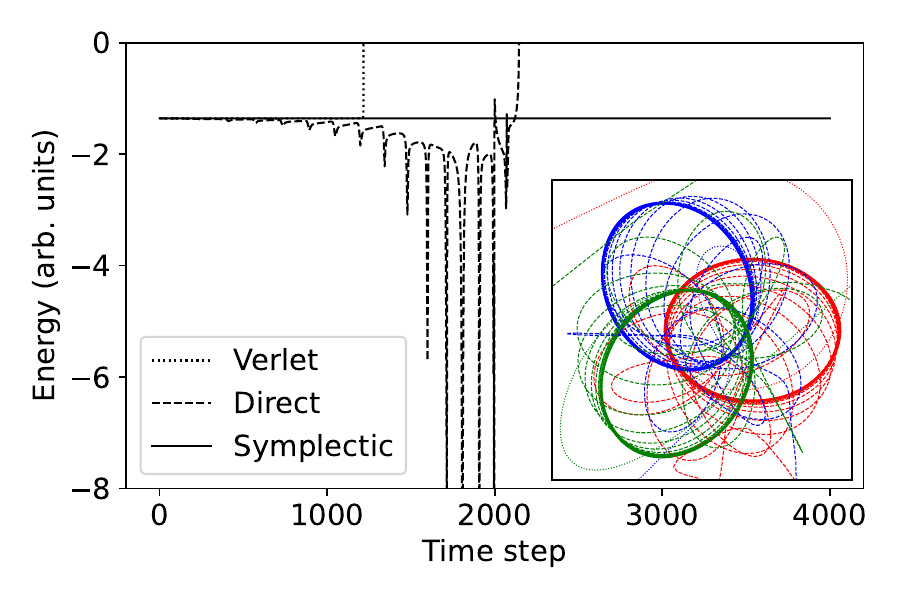}
    \caption{Energy profiles and trajectories of direct and symplectic methods for \rev{the} simulation \rev{of a symmetric three body problem} with large time steps. A velocity Verlet simulation with the same large time step is also shown.}
    \label{fig:3body}
\end{figure}

\begin{figure*}[tb]
    \centering
    \includegraphics[width=\linewidth]{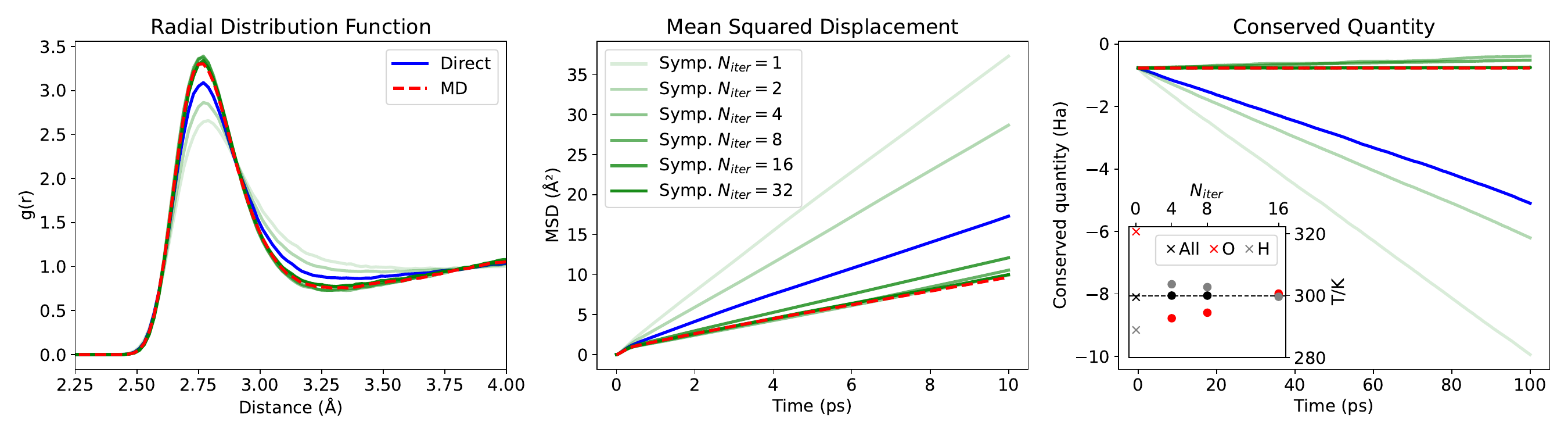}
    \caption{Simulations of liquid water performed in the NVT ensemble at 300 K, comparing a velocity Verlet baseline, direct predictions, and symplectic and time-reversible predictions using a variable number of fixed-point iterations per time step. Left: oxygen radial distribution function. Center: mean squared displacement of oxygen atoms. Right: profile of the conserved quantity (total energy plus themostat exchange energy), with an inset representing the average atom-type-resolved kinetic temperatures.}
    \label{fig:water}
\end{figure*}

In order to illustrate these benefits in practical simulations, we shall examine a few case studies. In all the examples that follow, we generate reference data by running short classical trajectories using  velocity Verlet~\cite{verl67pr} integration with a small time step. This integration method is ideal to generate the reference data, as it is explicit, symplectic and time-reversible for the systems we consider, since they all have a separable Hamiltonian $H(\bp, \bq) = T(\bp) + V(\bq)$. If this were not the case, generation of reference data using the implicit midpoint rule would be more appropriate.
We then train both a ``direct prediction'' model  which simply predicts the future coordinates after a large fixed time delay as a function of $(\bpbar, \bqbar)$, and a second model that instead predicts  $\Tilde{S}^3(\bpbar, \bqbar)$. 
We discuss the architecture, the hyperparameters and the training set construction for all the examples in the SI.

At training time, the evaluation of Eq.~\ref{eq:s3-predictions} poses no problems, since $(\bp, \bq)$, $(\bp', \bq')$ and hence $(\bpbar,\bqbar)$ are known\rev{, and the model can be trained to reproduce the reference $\Delta\bp$ and $\Delta\bq$ through Eq.~\ref{eq:s3-predictions}}. However, the prediction of the dynamics of the system (i.e., the prediction of $(\bp', \bq')$ from $(\bp, \bq)$ \rev{alone}) involves the solution of an implicit problem. Since the latter takes the exact form of the implicit midpoint rule, we use standard techniques from the integration of Hamiltonian systems, namely the use of fixed-point iterations~\cite{hairer2010geometric}, stabilized with a mixing approach, to solve the implicit system. We initialize the iteration with the direct prediction model, and then iterate until convergence, or perform a fixed number of iterations, which amounts to applying a correction to the non-structure-preserving model. \rev{In the SI, we provide further details on the method and we investigate the use of Anderson acceleration with finite memory to improve the convergence rate of the fixed-point iterations used to solve the implicit system.}

As a first illustrative example, we consider the prediction of the dynamics of a 3-body problem in a symmetric configuration that admits a closed solution with periodic orbits. 
The predictions of the trajectories of the three bodies by the two models, run with a large step for which velocity Verlet is unstable, are shown in Fig.~\ref{fig:3body}, together with  the corresponding total energy profiles. 
The results of the symplectic model, iterated to convergence, show the remarkable long-time stability and accuracy that is typical of symplectic integrators, which allow the definition of a modified Hamiltonian, while the direct model displays an unphysical precession and poor energy conservation along the trajectory.

While this three-body problem showcases the desirable properties of the proposed structure-preserving method, applications of classical dynamics often involve a much larger number of \rev{interacting particles}, and complicated many-body potential energy functions. This is the case in molecular dynamics, where the goal is often to obtain thermodynamic averages from microscopic simulations.

To illustrate atomistic applications, we consider the archetypal case of simulations of liquid water. 
We first consider a direct-prediction ML integrator based on a FlashMD architecture, trained on $NVE$ molecular dynamics trajectories of the solid and liquid phases across different  densities (between 90\% and 110\% of the experimental density of water) and temperatures (from \rev{1}0 to 1000 K). Molecular dynamics was performed using the q-TIP4P/f~\cite{habershon2009competing} potential from the i-PI simulation package~\cite{litm+24jcp}, using a \rev{velocity} Verlet integrator with a conservative time step of 0.25~fs. As in the previous example, the direct model predicts the future positions and momenta after a fixed time interval, which we combine with a stochastic velocity rescaling thermostat~\cite{buss+07jcp} to perform simulations in the constant-temperature ensemble, substituting the velocity Verlet step in a symmetric Trotter split integrator (the so-called OBABO integrator~\cite{buss-parr07pre}). 
When using a time step of 2~fs, the direct integrator shows a large drift of the conserved quantity, violation of energy equipartition, and noticeable deviations in the computed static and dynamical properties relative to the reference short-time-step MD results~(Figure~\ref{fig:water}). These are sampling artifacts that are common to all current ML integrators.
We then train a model for $S^3$ using a similar graph neural network architecture (see the SI for model details) and use it in an iterative way, starting from the direct prediction of $(\bp',\bq')$ and applying a prescribed number of fixed-point iterations.
As shown in Fig.~\ref{fig:water}, as the number of iteration is increased the integrator converges to be structure-preserving, progressively reducing energy drift and kinetic temperature imbalance between O and H atoms. 
The O-O pair correlation function and the mean-square displacement curves (reporting on the structural and diffusive properties of water) also converge to the reference values, providing a striking demonstration of the \rev{beneficial effects of enforcing} Hamiltonian structure in ML integrators.

\begin{figure}[tb]
    \centering
    \includegraphics[width=\linewidth]{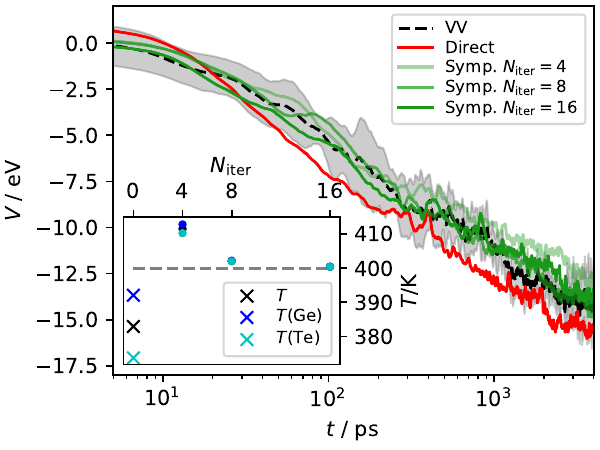}
\caption{Potential energy relaxation for long-time simulation of deeply-undercooled GeTe ($T=400$~K), using a cubic box containing 432 atoms. 
The different curves correspond to the reference velocity Verlet simulations (black), direct trajectory prediction (red) and symplectic corrections with different numbers of fixed-point iterations (shades of green). The curves are smoothed with a moving average with a Gaussian window of 2~ps, and averaged over 4 independent runs. The gray band indicates a range of two standard errors around the mean for the VV reference; error bars for the other curves are hidden, for clarity, but are of a similar magnitude. 
The inset shows the mean temperature (black) as well as the temperature resolved between Ge (blue) and Te (cyan), for direct predictions ($N_{\text{iter}}=0$) and for different levels of symplectic iterations.
}
    \label{fig:gete}
\end{figure}

As an even more challenging example, we consider the case of the phase-change material GeTe. We take inspiration from the simulations in Ref.~\cite{soss+13jpcl} and \rev{investigate} this system in the deep undercooling regime, where it exhibits glassy behavior.
In particular, we observe that in the reference MD simulations the potential energy relaxes with a logarithmic behavior that persists up to a time scale of several ns. 
We train direct and symplectic models based on short trajectories launched from configurations collected along a constant-pressure temperature cycle in which we raised the temperature from 100 to 1500~K over 400~ps, and then quenched back to 100~K, using the PET-MAD  universal interatomic potential~\cite{mazitov2025pet}. 
We then run ns-scale trajectories, held at 400~K with a gentle stochastic velocity rescaling thermostat~\cite{buss+07jcp}, both with a velocity Verlet integrator with a 4~fs time step, and with ML-based integrators with a much longer, 30~fs time step. Despite the limited amount of training data, the direct model shows only small deviations from the target temperature, and equipartition between Ge and Te atoms is broken by less than 20~K. A few iterations of the symplectic corrections enforce equipartition and reduce the error in the kinetic temperature, with $N_\text{iter}=16$ being sufficient to equilibrate fully. 
The good performance of the ML integrators is also reflected in their ability to capture quantitatively the glass-like relaxation. Even the direct prediction of the trajectory is barely outside the confidence region of the reference simulations, and all the runs using symplectic corrections are within the statistical error bars. 

\rev{An even more complete set of} examples, including the use of larger time steps, NVE and NPT simulations and different \rev{baseline} potentials are \rev{ also} discussed in the SI\rev{, where we also show preliminary evidence that this approach can be applied across different material compositions, with a universal model trained on a general-purpose dataset, which we then test on six highly diverse systems.} Collectively, the\rev{se experiments} indicate that violation of structure-preservation properties is the main shortcomings of emerging ML approaches to extend the integration intervals for classical dynamics in general, and molecular simulations in particular.
The violation of long-time energy conservation and equipartition are good indirect diagnostics for the severity of the problem.
We have proposed a practical method to parametrize structure-preserving maps, which dramatically improves the stability of ML integrators for a given time step, and cures energy conservation and equipartition issues, recovering the equilibrium and, perhaps more impressively, dynamical properties of conventional small-time-step Hamiltonian integrators for realistic simulation problems. 

A current limitation of our approach is the need to evaluate derivatives by back-propagation, and to use an implicit mid-point integrator -- both increasing substantially the computational cost over a direct prediction \rev{(see Fig. S3 in the SI for some representative examples).}
The possibility of applying the iterations of the implicit solver as a correction -- that we show to improve systematically the accuracy towards the converged integrator -- provides a mitigation strategy.
From a machine-learning perspective, future work could focus on improving the accuracy of the direct predictions, using the symplectic version as a sanity check and as a practical way to monitor the violation of the Hamiltonian geometric structure, which would otherwise require an exceedingly expensive calculation of the Jacobian of the map.
On a more general level, the fact that we learn a generating function that is, modulo a trivial transformation, equivalent to the Hamilton-Jacobi action, suggests that similar ideas may be useful to approximate in a data-driven fashion the long-time-step dynamics of relativistic, quantum mechanical, and other types of systems and physical theories which can be formulated in terms of an action.

\section{Data availability}

The code, models and datasets to support this work can be found at \texttt{zenodo.org/records/16274506}.

\begin{acknowledgments}
\end{acknowledgments}

\bibliographystyle{unsrtnat}

\clearpage
\setcounter{secnumdepth}{2} 
\setcounter{section}{0}     
\setcounter{figure}{0}      
\setcounter{table}{0}       
\setcounter{equation}{0}    

\renewcommand{\thesection}{S\arabic{section}}
\renewcommand{\thesubsection}{\thesection.\arabic{subsection}}
\renewcommand{\thefigure}{S\arabic{figure}}
\renewcommand{\thetable}{S\arabic{table}}
\renewcommand{\theequation}{S\arabic{equation}}

\makeatletter
\def\@hangfrom@section#1#2#3{\@hangfrom{#1#2\quad}#3}
\makeatother

\section{Generating functions of symplectic maps}
\label{si:generating-functions}

A symplectic map $\mathcal{S}: (\bp, \bq) \rightarrow (\bp', \bq')$ is a differentiable map which possesses the following property:
\begin{equation}
    \Big(\frac{d\boldsymbol{y}'}{d\boldsymbol{y}}\Big)^\top \boldsymbol{J} \,\, \Big(\frac{d\boldsymbol{y}'}{d\boldsymbol{y}}\Big) = \boldsymbol{J},
\end{equation}
where $\boldsymbol{y} \equiv (\bp, \bq)$, $\boldsymbol{y}' \equiv (\bp', \bq')$, $d\boldsymbol{y}'/d\boldsymbol{y}$ is the Jacobian matrix of the map, and
\begin{equation}
    \boldsymbol{J} =
    \begin{pmatrix}
        \boldsymbol{0} & \boldsymbol{I} \\
        -\boldsymbol{I} & \boldsymbol{0}
    \end{pmatrix}.
\end{equation}

In general, any symplectic map can be associated with a generating function which fully characterizes it. The generating function is a scalar function that can be expressed in a number of ways. \rev{T}hose mentioned in the main text \rev{are}:
\begin{equation}
    S(\bq, \bq'), \quad \bp = - \frac{\partial S}{\partial \bq}, \quad \bp' = \frac{\partial S}{\partial \bq'},
\end{equation}
\begin{equation}
    S^1(\bp', \bq), \quad \Delta \bq = \frac{\partial S^1}{\partial \bp'}, \quad \Delta \bp = - \frac{\partial S^1}{\partial \bq},
\end{equation}
\begin{equation}
    S^2(\bp, \bq'), \quad \Delta \bq = \frac{\partial S^2}{\partial \bp}, \quad \Delta \bp = - \frac{\partial S^2}{\partial \bq'},
\end{equation}
\begin{equation}
    S^3(\bpbar, \bqbar), \quad \Delta \bq = \frac{\partial S^3}{\partial \bpbar}, \quad \Delta \bp = - \frac{\partial S^3}{\partial \bqbar}.
    \label{eq:s3}
\end{equation}

\section{The choice of generating functions of the third type}\label{si:third-type}

Although, in principle, all the \rev{generating functions} highlighted in \ref{si:generating-functions} can be represented by a neural network and learned, the generating function of the third type $S^3$ is particularly \rev{convenient}.

First, we consider the problem of enforcing time-reversibility on the symplectic map and, therefore, on the generating function. Time-reversibility can be expressed as the statement that, if the symplectic map $\mathcal{S}$ gives $(\bp, \bq) \rightarrow (\bp', \bq')$, then it must also give $(-\bp', \bq') \rightarrow (-\bp, \bq)$. From the equations in \ref{si:generating-functions}, one can see that this implies (up to additive constants):
\begin{equation}
    S(\bq, \bq') = S(\bq', \bq),
\end{equation}
\begin{equation}
    S^1(\bp', \bq) = S(-\bp, \bq'),
\end{equation}
\begin{equation}
    S^2(\bp, \bq') = S^2(-\bp', \bq),
\end{equation}
\begin{equation}
    S^3(\bpbar, \bqbar) = S^3(-\bpbar, \bqbar).
\end{equation}
The conditions for $S^1$ and $S^2$ would lead to complicated implementations, since the targets of the machine learning exercise would also need to be used as inputs to enforce time-reversibility.

\rev{Moreover}, the choice of $S(\bq, \bq')$, while practical at training time, poses problems at inference time: one would have to find $\bq'$ so as to obtain a self-consistent prediction of $\bp$, which is already known. However, solving this problem numerically leads to two slightly different values of $\bp$: one known from the previous step and one given by the generating function $S$ calculated at the current step. This is not the case if one chooses $S^3(\bpbar, \bqbar)$, where finite precision does not lead to potential changes in the values of the positions and momenta at the current step, but only affects those at the future step, as in traditional implicit integrators. Furthermore, the expressions to propagate the generating function of the third type $S^3$ are exactly equivalent to those for the implicit midpoint rule, making it possible to use established and optimized implementations for this traditional implicit integrator.

\rev{
Finally, $S(\bq, \bq', h)$ features a caustic at $h=0$ which could affect training and prediction stability, while the first caustic of $S^3(\bpbar, \bqbar, h)$ is delayed (for example, it is located at half the period of oscillation in a harmonic oscillator). For more details, we redirect the reader to the discussion below related to the 16 fs predictions on the liquid water system by the universal molecular dynamics models.
}

\section{Relationship to traditional integrators}\label{si:traditional integrators}

The relationship of our method to traditional integrators, and in particular to the implicit midpoint rule, can provide additional insight into how the proposed parametrization of symplectic maps works. The implicit midpoint rule is based on the truncation of $S^3(\bpbar, \bqbar, h)$ to its leading-order term in the step size $h$~\cite{hairer2010geometric}:
\begin{equation}\label{eq:s3_for_midpoint_integrator}
    S^3(\bpbar, \bqbar, h) \approx hH(\bpbar, \bqbar).
\end{equation}
While this approximation is good for small $h$, it is not accurate for large time steps. By providing an arbitrary parametrization of $S^3$, our method seeks instead to represent it exactly, and it can therefore afford physically faithful dynamics using larger time steps.
\rev{
As a pedagogical example, the $S^3$ generating function for a one-dimensional harmonic oscillator is
\begin{equation}\label{eq:s3_for_harmonic_oscillator}
    S^3(\Bar{p}, \Bar{q}, h) = \frac{2}{\omega} \, \textrm{tan} \Big( \frac{\omega h}{2} \Big) \, H(\Bar{p}, \Bar{q}),
\end{equation}
where $\omega$ is its angular frequency and $H(\Bar{p}, \Bar{q}) = \Bar{p}^2 / (2m) + m\omega^2 \Bar{q}^2 / 2$ is its Hamiltonian evaluated at the midpoint variables. A natural consequence is that learning the Hamiltonian $H(\bp, \bq)$~\cite{greydanus2019hamiltonian} instead of a symplectic map would not eliminate the need for a small-time-step integrator, as propagation would rely on the approximation in Eq.~\ref{eq:s3_for_harmonic_oscillator}, which is in general only valid for small time steps $h$.
}

\section{Small deviations from equipartition in the momentum terms of classical Hamiltonians}\label{si:small-deviations-from-equipartition}

Just like traditional numerical integrators, our method is not guaranteed to afford exact equipartition of energy in its most intuitive form. Here we discuss the reasons for this and we analyze an expansion in the shadow Hamiltonian that can provide valuable insights in this regard.

Let the shadow Hamiltonian $H(\bp, \bq)$ be the solution of the Hamilton-Jacobi equation with $S$ defined by the neural network (possibly via $S^3$). Then, the equipartition theorem holds in the canonical ensemble (see, e.g., Chapter 3 of Ref.~\cite{pathria}) in the sense that
\begin{equation}\label{eq:equipartition}
    \left\langle x_m\frac{\partial H}{\partial x_n} \right\rangle = \delta_{mn} k_B T,
\end{equation}
where $x_m$ and $x_n$ are individual components of the position or momentum vectors $\bq$ and $\bp$. However, since the shadow hamiltonian $H$ does not, in general, take the form 
\begin{equation}\label{eq:hamiltonian-si}
    H(\bp, \bq) = \sum_i \frac{p_i^2}{2m_i} + V(\bq),
\end{equation}
certain common and intuitive consequences of the equipartition theorem are not followed. For example, with $H$ as in Eq.~\ref{eq:hamiltonian-si}, Eq.~\ref{eq:equipartition} implies
\begin{equation}
    \frac{\langle p^2 \rangle}{2m} = \frac{1}{2} k_B T,
\end{equation}
where $p$ can be any component of $\bp$ and $m$ is the mass relative to the corresponding degree of freedom.

This is not necessarily the case for a general shadow Hamiltonian. To see why, we expand $H(\bp, \bq)$ around $\bp = \boldsymbol{0}$ and we evaluate $\langle p^2 \rangle$:
\begin{equation}\label{eq:intuitive-equipartition}
    \langle p^2 \rangle = \frac{1}{Z} \int \Bigg( \int p^2 e^{H(\bp, \bq) / k_B T} d\bp \Bigg) d\bq,
\end{equation}
\begin{equation}
    H(\bp, \bq) = c_0 (\bq) + \boldsymbol{c}_1(\bq)^\top \bp + \bp^\top \boldsymbol{C}_2(\bq) \, \bp + ...
\end{equation}
One can see that $c_0(\bq)$ has no effect on the value of the integral (as it simplifies with the corresponding term in $Z$, exactly like the $V(\bq)$ term in a Hamiltonian in the form of Eq.~\ref{eq:hamiltonian-si}). The term $\boldsymbol{c}_1(\bq)^\top \bp$ does have an effect, but it vanishes if time-reversible dynamics are enforced (due to the condition $H(-\bp, \bq) = H(\bp, \bq)$), together with the cubic term in $\bp$ and all other odd terms. Now, for Eq.~\ref{eq:intuitive-equipartition} to hold for all momentum degrees of freedom, we would need $\boldsymbol{C}_2$ to be diagonal, independent of $\bq$, with entries corresponding to the inverse of twice the mass of each respective degree of freedom, and, finally, all higher-order even terms would need to be neglected. While these conditions are not true in general, accurate models of the action would satisfy them approximately.

This analysis highlights the importance of enforcing symplecticity (which allows the definition of the shadow Hamiltonian in the first place), as well as time-reversibility (which eliminates the most problematic terms in its expansion), in order to obtain models that better satisfy intuitive equipartition of the energy associated with quadratic terms in the momentum from the original Hamiltonian of the system.

\section{One-body orbit}\label{si:one-body}

\begin{figure}
    \centering
    \includegraphics[width=0.7\linewidth]{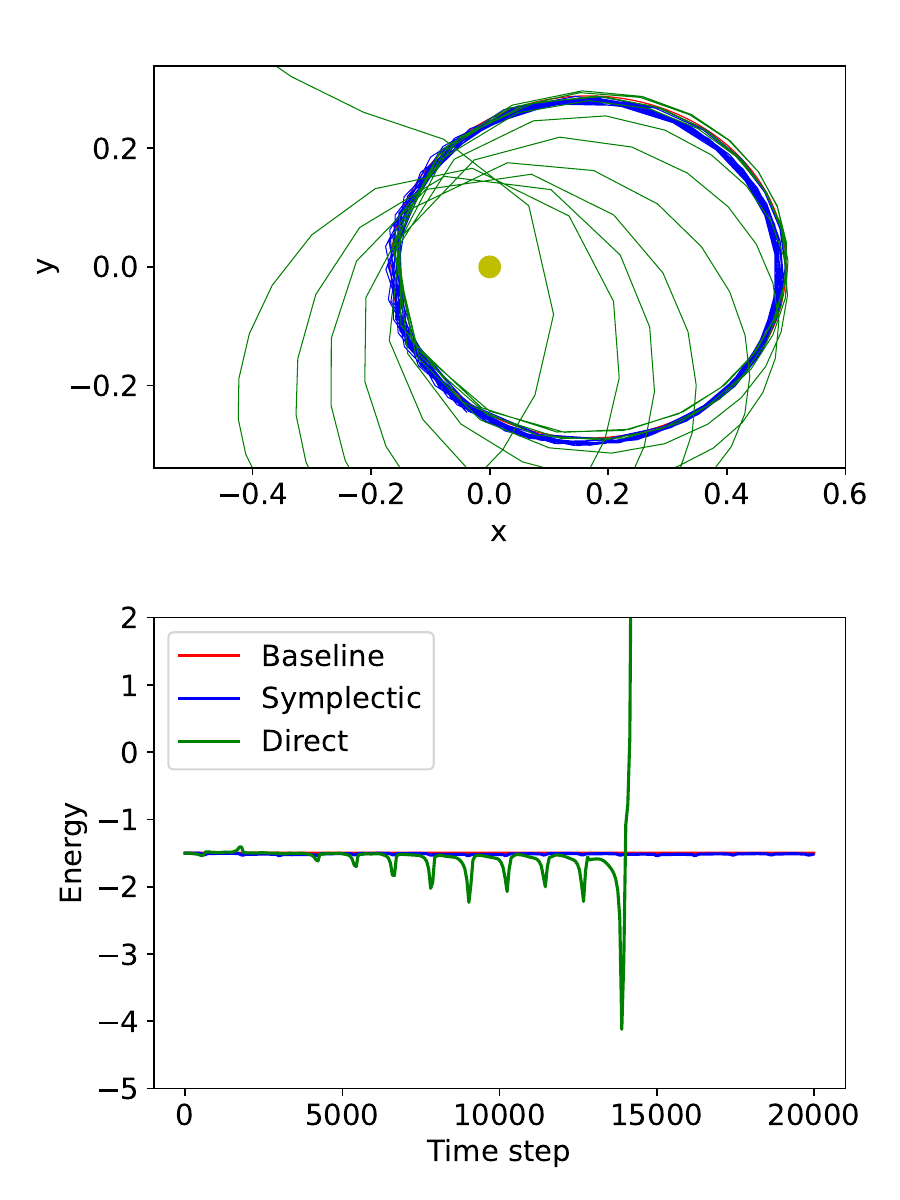}
    \caption{Trajectories (top) and total energy profiles (bottom) for a one-body simulation for (1) a baseline trajectory performed using velocity Verlet with a small time step, (2) a symplectic and time-reversible trajectory predictor, and (3) a direct predictor.}
    \label{fig:1body}
\end{figure}

Since the set-up for the three-body problem presented in the main text can be tedious to reproduce, here we present a version where the favorable properties of a structure-preserving dynamics predictor are showcased in a much simpler setting. This example concerns an orbit simulation of a single body around a stationary mass. The machine-learning exercise is extremely simple, as the models for the time-evolution of the system only take the coordinates and momenta of one particle as inputs. Fig.~\ref{fig:1body} shows that, while the direct prediction method does not conserve energy and eventually diverges from the correct trajectory, the structure-preserving method conserves energy and yields a \rev{stable} trajectory over long times.

\section{Molecular dynamics of Lennard-Jones argon}\label{si:argon}

The molecular dynamics examples exposed in the main text are relatively involved and/or use potentials that might not be readily accessible from the most popular molecular dynamics simulation engines. Here, we present a molecular dynamics example which is particularly simple, as it uses a standard potential (Lennard-Jones) from a widely available simulation package (LAMMPS~\cite{thompson2022lammps}). Furthermore, compared to the systems investigated in the main text, it only contains one chemical species, it does not make use of advanced sampling techniques, and it involves training on trajectories generated for a single thermodynamic state point. While the main text examples involve dynamics in the NVT ensemble, here we will restrict ourselves to the NVE ensemble, which is easier to implement in exploratory codes. We believe that this molecular dynamics example can serve as a minimalistic case study that can be reproduced with ease, while still showcasing the advantages of symplectic trajectory prediction very clearly.

\begin{figure}
    \centering
    \includegraphics[width=0.7\linewidth]{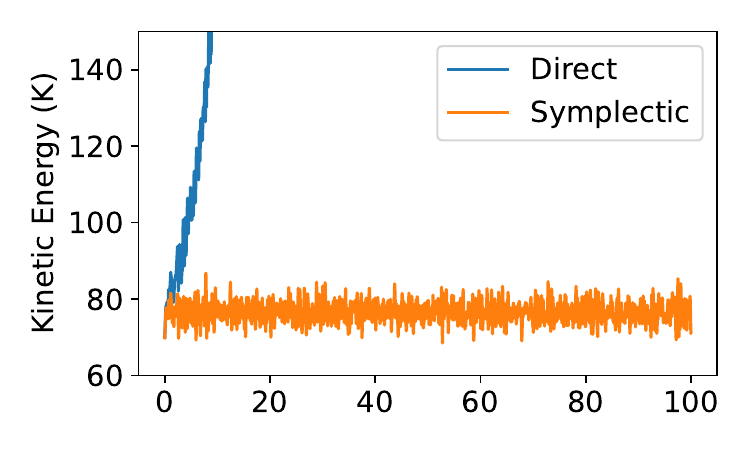}
    \caption{Kinetic energy (expressed in K), of two NVE simulations of liquid argon performed with a direct prediction model and with a symplectic model.}
    \label{fig:argon}
\end{figure}

We train two models on NVE molecular dynamics trajectories of liquid argon using a Lennard-Jones potential. As in the previous example, one neural network predicts the future positions and momenta after a time interval of 16 fs directly, while the other does so using a parametrizable symplectic and time-reversible map. Fig.~\ref{fig:argon} shows that the symplectic approach affords excellent energy conservation, while the direct model rapidly heats the system in an unphysical manner.

\section{A case study in the NPT ensemble}\label{si:npt}

In order to explore the feasibility of dynamics in the NPT ensemble (in addition to the NVT ensemble presented in the main text and the NVE ensemble used in \ref{si:argon}), we use a symplectic molecular dynamics predictor to simulate solid-state face-centered-cubic aluminum at 300 K.

We perform simulations of aluminum with the PET-MAD universal interatomic potential and we train a symplectic neural network to predict molecular dynamics trajectories, both in the solid and liquid states, creating a transferable molecular dynamics model for bulk aluminum. To test the physical correctness of the model, we perform NPT simulations, recording the average volume, temperature, and the time taken to perform a 100 ps simulation, both with the PET-MAD machine-learned universal interatomic potential and with the symplectic model for molecular dynamics prediction. These are reported in Table~\ref{tab:aluminum}, showing that the symplectic model reproduces observables correctly while affording a speedup over a simple molecular dynamics simulation with a machine-learned interatomic potential.
\begin{table}[h]
    \centering
    \begin{tabular}{l@{\hspace{10pt}}c@{\hspace{7pt}}c@{\hspace{7pt}}c}
    \toprule
         Simulation & T (K) & V (\AA$^3$) & Time (h) \\
    \midrule
         MLIP & $300.5 \pm 0.9$ & $1782.1 \pm 0.0$ & 1.389 \\
         Symplectic NN & $301.5 \pm 1.2$ & $1783.0 \pm 0.2$ & 0.227 \\
    \bottomrule
    \end{tabular}
    \caption{Kinetic temperatures, volumes and timings for 100 ps NPT simulations of aluminum, using a machine-learned interatomic potential (MLIP) and a symplectic model. Timings are measured on a Nvidia H100 GPU.}
    \label{tab:aluminum}
\end{table}

The proposed method, while achieving correct sampling of the desired thermodynamic ensemble, does not afford any acceleration when compared against empirical potentials such as those used for water and argon, as these are very cheap to compute compared to neural networks. However, as shown in this example, the proposed symplectic method can achieve a speedup over neural-network-based potentials, which are increasingly used in atomistic modeling.

\section{Monitoring molecular dynamics quality with symplectic predictors}

Especially when using symplectic integrators in the context of molecular dynamics simulations, we found that the quality of the simulation correlates strongly with its degree of energy conservation. This can be understood in terms of the sampling of thermodynamic ensembles in molecular dynamics, which relies on two properties being satisfied by the discrete integrator:
\begin{itemize}
    \item Liouville's theorem, which states that the Hamiltonian flows preserves phase-space volume, including its orientation
    \item Conservation of energy, which is satisfied exactly by Hamiltonian flow
\end{itemize}
While all symplectic methods satisfy the first condition, the second is satisfied only approximately during traditional numerical integration, and it is therefore crucial to monitor it during a simulation. The probability of observing a given microstate $(\bp, \bq)$ is
\begin{equation}
    \frac{1}{Z} P(H(\bp, \bq)) \, d\bp \, d\bq,
\end{equation}
where $P$ is a probability density function, which depends on the Hamiltonian (and thermodynamic quantities) and changes for different thermodynamic ensembles, and $Z$ is the partition function. If the two conditions above are satisfied, it follows that
\begin{itemize}
    \item $d\bp' \, d\bq' = d\bp \, d\bq$, from the first condition,
    \item $H(\bp', \bq') = H(\bp, \bq)$, from the second condition.
\end{itemize}
Given these two equalities, the step of discretized dynamics preserves the probability of observing a given microstate. As a result, it preserves the probability distribution of any thermodynamic ensemble.

The quality of sampling of a symplectic method is therefore directly related to its degree of energy conservation. This is indeed what we observe in our experiments, where, when varying the degree of convergence of the fixed-point iterations, monitoring energy conservation is a very good proxy for the quality of the physical observables which can be extracted from the dynamics.

\section{Model, training and simulation details}\label{si:models-and-training}

In this section, we present the methods that were used to train the models used in this study and to perform simulations with them. For more details, the reader can refer to the code at \texttt{zenodo.org/records/16274506}.

\subsection{Models}

\subsubsection{One-body and three-body simulations}

For these simulations, we used simple multi-layer perceptrons. In the one-body case, we used a simple multi-layer perceptron with SiLU activation functions and sizes [4, 128, 128, 4] for the direct prediction of positions and momenta, and sizes [4, 128, 128, 1] for the prediction of the symplectic map. The same architecture was used to learn the dynamics of the three-body case, with the shapes of the multi-layer perceptrons changing to [12, 128, 128, 12] for direct predictions and [12, 128, 128, 1] for structure-preserving predictions. Structure-preserving predictions were additionally symmetrized as explained in the main text in order to enforce time-reversibility. 

\subsubsection{Molecular dynamics}

For models to be used in molecular dynamics, we used the FlashMD architecture (a graph neural network), which was used exactly in the same way as in the original paper~\cite{bigi2025flashmd} for direct molecular dynamics predictions, and modified to predict a single scalar in the case of structure-preserving predictions, in which case the predictions were additionally symmetrized to enforce time-reversibility.

\subsection{Data generation}

\subsubsection{One-body problem}

We generate data to learn the dynamics for this problem by simulating the orbit of a single body around a fixed stationary mass using the velocity Verlet algorithm with a time step of 0.001 for 100\,000 steps. All masses, as well as the gravitational constant $G$, are set to 1. The initial conditions are set to $q_x = 1/2, q_y = 1, p_x = 0, p_y = 1$. All possible $((\bq, \bp), (\bq', \bp'))$ pairs from these trajectories, with a time separation of 64 velocity Verlet steps, were included in the training set.

\subsubsection{Three-body problem}

We generate data to learn the dynamics for this problem by simulating the motion of three bodies using the velocity Verlet algorithm with a time step of 0.0001 for 200\,000 steps. All masses, as well as the gravitational constant $G$, are set to 1. The initial conditions are set to $q_{1,x} = 1, q_{1,y} = 0, q_{2,x} = -1/2, q_{2,y} = \sqrt{3}/2, q_{3,x} = -1/2, q_{3,y} = -\sqrt{3}/2, p_{1,x} = 0, p_{1,y} = 1/2, p_{2,x} = -\sqrt{3}/4, p_{2,y} = -1/4, p_{3,x} = \sqrt{3}/4, p_{3,y} = -1/4$.  All possible $((\bq, \bp), (\bq', \bp'))$ pairs from these trajectories, with a time separation of 256 velocity Verlet steps, were included in the training set.

\subsubsection{Water}

We take a structure that was equilibrated using the q-TIP4P/f potential~\cite{habershon2009competing} from i-PI~\cite{litm+24jcp} at 300 K in the NVT ensemble and using the experimental density of water at room temperature and pressure. From this structure, we scale all positions and cell coordinates to generate two more structures with cell volumes reduced and augmented by 10\%, respectively. For these three structures, and for all temperatures going from 20 K to 1000 K (both included) in steps of 20 K, we run an equilibration trajectory in the NVT ensemble for 25 ps using a Langevin thermostat~\cite{buss-parr07pre} with a time constant of 10 fs and a step size of 0.5 fs. Subsequently, for each equilibrated structure, we perform simulations in the NVE ensemble, using the velocity Verlet integration algorithm~\cite{verl67pr}, for 2 ps using a time step of 0.25 fs. All trajectories run in the NVE ensemble are included in the training set, taking $(\bq, \bp)$ pairs at intervals of 100 fs and the corresponding $(\bq', \bp')$ pairs 2 fs (or, equivalently, 8 velocity Verlet steps) after $(\bq, \bp)$. The dataset for the non-structure-preserving method is augmented with the time-reversed version of each $((\bq, \bp), (\bq', \bp'))$ training sample.

\subsubsection{GeTe}

We take a GeTe structure from a publicly available database~\cite{dang+21npjcm}, containing 512 atoms in total. From this structure, we run two constant-pressure simulations, using the barostat from Ref.~\cite{bussi2009isothermal}: one where the temperature is increased linearly, from 100 K to 1500 K, over a duration of 400 ps, using a stochastic velocity rescaling~\cite{buss+07jcp} thermostat, with a time constant of 10 fs and a step size of 4 fs, and a trajectory starting from the last structure of the first, where the temperature is instead decreased from 1500 K to 100 K over the same time span. Structures are then collected, along both paths, at simulation times corresponding to temperatures ranging from 200 K to 1400 K (both included) in steps of 25 K. For each of these structures, we run an equilibration trajectory in the NVT ensemble at the respective temperature for 50 ps using a Langevin thermostat~\cite{buss-parr07pre} with a time constant of 10 fs and a step size of 1 fs. Subsequently, for each equilibrated structure, we perform simulations in the NVE ensemble, using the velocity Verlet integration algorithm~\cite{verl67pr}, for 8 ps using a time step of 1 fs. All trajectories run in the NVE ensemble are included in the training set, taking $(\bq, \bp)$ pairs at intervals of 1000 fs and the corresponding $(\bq', \bp')$ pairs 30 fs (or, equivalently, 30 velocity Verlet steps) after $(\bq, \bp)$. The dataset for the non-structure-preserving method is augmented with the time-reversed version of each $((\bq, \bp), (\bq', \bp'))$ training sample.

\subsubsection{Argon}

We simulate liquid Argon using a Lennard-Jones potential, with parameters $\epsilon = 0.0103$ eV, $\sigma = 3.4$~\AA, and an interaction cutoff of 10~\AA, using the LAMMPS simulation package~\cite{thompson2022lammps}. Ten simulations are initialized using a 4x4x4 supercell, where each individual face-centered-cubic cell has an edge length of 4.05~\AA, and where the velocities are sampled from a Maxwell-Boltzmann distribution at 80 K. For each simulation, an equilibration run in the NPT ensemble is performed for 100 ps using a step size of 1 fs, followed by an NVE trajectory using the velocity Verlet algorithm~\cite{verl67pr} for 10 ps using a step size of 1 fs. All trajectories run in the NVE ensemble are included in the training set, taking $(\bq, \bp)$ pairs at intervals of 400 fs and the corresponding $(\bq', \bp')$ pairs 16 fs (or, equivalently, 16 velocity Verlet steps) after $(\bq, \bp)$. The dataset for the non-structure-preserving method is augmented with the time-reversed version of each $((\bq, \bp), (\bq', \bp'))$ training sample.

\subsubsection{Aluminum}

We take a face-centered-cubic aluminum structure with experimental density at room temperature and pressure. From this structure, we run two constant-pressure simulations, using the barostat from Ref.~\cite{bussi2009isothermal}: one where the temperature is increased linearly, from 100 K to 1500 K, over a duration of 1 ns, using a stochastic velocity rescaling~\cite{buss+07jcp} thermostat, with a step size of 2 fs, and a similar trajectory where the temperature is instead decreased from 1500 K to 100 K over the same time span. Structures are then collected, along both paths, at simulation times corresponding to temperatures ranging from 200 K to 1400 K (both included) in steps of 25 K. For each of these structures, we run an equilibration trajectory in the NVT ensemble at the respective temperature for 50 ps using a Langevin thermostat~\cite{buss-parr07pre} with a time constant of 10 fs and a step size of 1 fs. Subsequently, for each equilibrated structure, we perform simulations in the NVE ensemble, using the velocity Verlet integration algorithm~\cite{verl67pr}, for 8 ps using a time step of 1 fs. All trajectories run in the NVE ensemble are included in the training set, taking $(\bq, \bp)$ pairs at intervals of 400 fs and the corresponding $(\bq', \bp')$ pairs 30 fs (or, equivalently, 30 velocity Verlet steps) after $(\bq, \bp)$. The dataset for the non-structure-preserving method is augmented with the time-reversed version of each $((\bq, \bp), (\bq', \bp'))$ training sample.

\subsection{Training}

\subsubsection{One-body and three-body problems}

Training is performed with the Adam optimizer~\cite{kingma2014adam} for 20 epochs, with a batch size of 8, an initial learning rate of 0.001 and a learning rate decrease of a factor of 0.7 every 10000 training steps. Rotational augmentation is used during training, meaning that, at every training step, a different random rotation is applied to each training sample in the batch.

\subsubsection{Molecular dynamics}

Training is performed using the \texttt{metatrain} package, with the Adam optimizer~\cite{kingma2014adam} for 800 epochs, with an initial learning rate of $1 \cdot 10^{-4}$ and a learning rate decrease of a factor of 0.5 upon stagnation of the validation metric for 100 epochs. Rotational and inversion augmentation is used during training, meaning that, at every training step, a different random (possibly improper) rotation is applied to each training sample in the batch.

\subsection{Simulations}

\subsubsection{One-body problem}

Simulations with the structure-preserving and non-structure-preserving models are run for 312 steps, with a step size of 0.064. The initial conditions are set to $q_x = 1/2, q_y = 1, p_x = 0, p_y = 1$.

\subsubsection{Three-body problem}

Simulations with the structure-preserving and non-structure-preserving models, as well as with a large-step velocity Verlet algorithm, are run for 4000 steps, with a step size of 0.0256. The initial conditions are set to $q_{1,x} = 1, q_{1,y} = 0, q_{2,x} = -1/2, q_{2,y} = \sqrt{3}/2, q_{3,x} = -1/2, q_{3,y} = -\sqrt{3}/2, p_{1,x} = 0, p_{1,y} = 1/2, p_{2,x} = -\sqrt{3}/4, p_{2,y} = -1/4, p_{3,x} = \sqrt{3}/4, p_{3,y} = -1/4$, one of the several known periodic solutions of the three-body problem. 

\subsubsection{Water}

Simulations with the structure-preserving and non-structure-preserving models are run for 100 ps, with a step size of 2 fs. The reference MD simulation is run with the q-TIP4P/f potential~\cite{habershon2009competing} for 100 ps and with a step size of 0.25 fs. All simulations are performed in the NVT ensemble at 300 K, using a stochastic velocity rescaling thermostat~\cite{buss+07jcp} with a time constant of 10 fs.

\subsubsection{GeTe}

Following a protocol similar to that in Ref.~\citenum{soss+13jpcl}, we first perform a long simulation of the liquid phase at 1000~K using a stochastic velocity rescaling thermostat with a time constant of 1~fs, extracting uncorrelated samples that are used as the starting point of 4 independent simulations. 
For each quench to 400~K, running for 20~ps with a thermostat time constant of 10~fs, and take the final configuration as the starting point of 5 trajectories, all weakly thermostatted with a stochastic velocity rescaling time constant of 1~ps: (1) a 4~ns conventional MD simulation using a velocity Verlet integrator, with a time step of 4~fs, which we use as the reference; (2) a 6-ns direct ML prediction trajectory with a time step of 30~fs; (3-5) 6-ns structure-preserving ML integrator runs, using 4, 8, 16 fixed-point iterations, with a mixing parameter of 0.3.

\subsubsection{Argon}

Simulations with the structure-preserving and non-structure-preserving models are run for 100 ps, with a step size of 16 fs, without the application of thermostats or barostats.

\subsubsection{Aluminum}

Simulations with the structure-preserving and non-structure-preserving models are run for 100 ps, with a step size of 30 fs. The reference MD simulation is run with the PET-MAD interatomic potential~\cite{mazitov2025pet} for 100 ps and with a step size of 1 fs. All simulations are performed in the NPT ensemble at 300 K and 1 bar, using a stochastic velocity rescaling thermostat~\cite{buss+07jcp} with a time constant of 100 fs and the barostat from Ref.~\cite{bussi2009isothermal} with a time constant of 1 ps, and where the cell degrees of freedom are coupled to a Langevin thermostat with a time constant of 1 ps.

\rev{
\subsection{Universal models}

\subsubsection{Overview}

In order to demonstrate the wide applicability of the symplectic construction described in this work, we train two universal models for the prediction of molecular dynamics across the periodic table: the first is trained to predict time steps of 2 fs, the second to predict time steps of 16 fs. We train both models on a dataset of molecular dynamics trajectories published as part of Ref.~\cite{bigi2025flashmd}, which were generated using the PET-OMATPES-L universal r2SCAN machine-learned force field~\cite{bigi2026pushing}, which was itself trained on the OMat24~\cite{barroso_omat24} and MATPES~\cite{kaplan2025foundational} datasets. We compare these models against published FlashMD models which predict the same step sizes but directly, i.e., without enforcing symplecticity, and which were trained on the same trajectory dataset. We perform the experiments on a wide range of chemical systems: bulk aluminum (Al), water (\ce{H2O}), gallium arsenide (\ce{GaAs}), barium titanate (\ce{BaTiO3}), lithium thiophosphate (\ce{Li3PS4}), and a high-entropy alloys (HEAs) with high chemical diversity. In addition to a bulk, we also ran a simulation showing the capabilities of our model on a slab. The results in fig.~\ref{fig:universal} show that a single model for symplectic dynamics can be used successfully across diverse systems to afford symplectic and energy-conserving dynamics.

\begin{figure*}
\centering
\includegraphics[width=\linewidth]{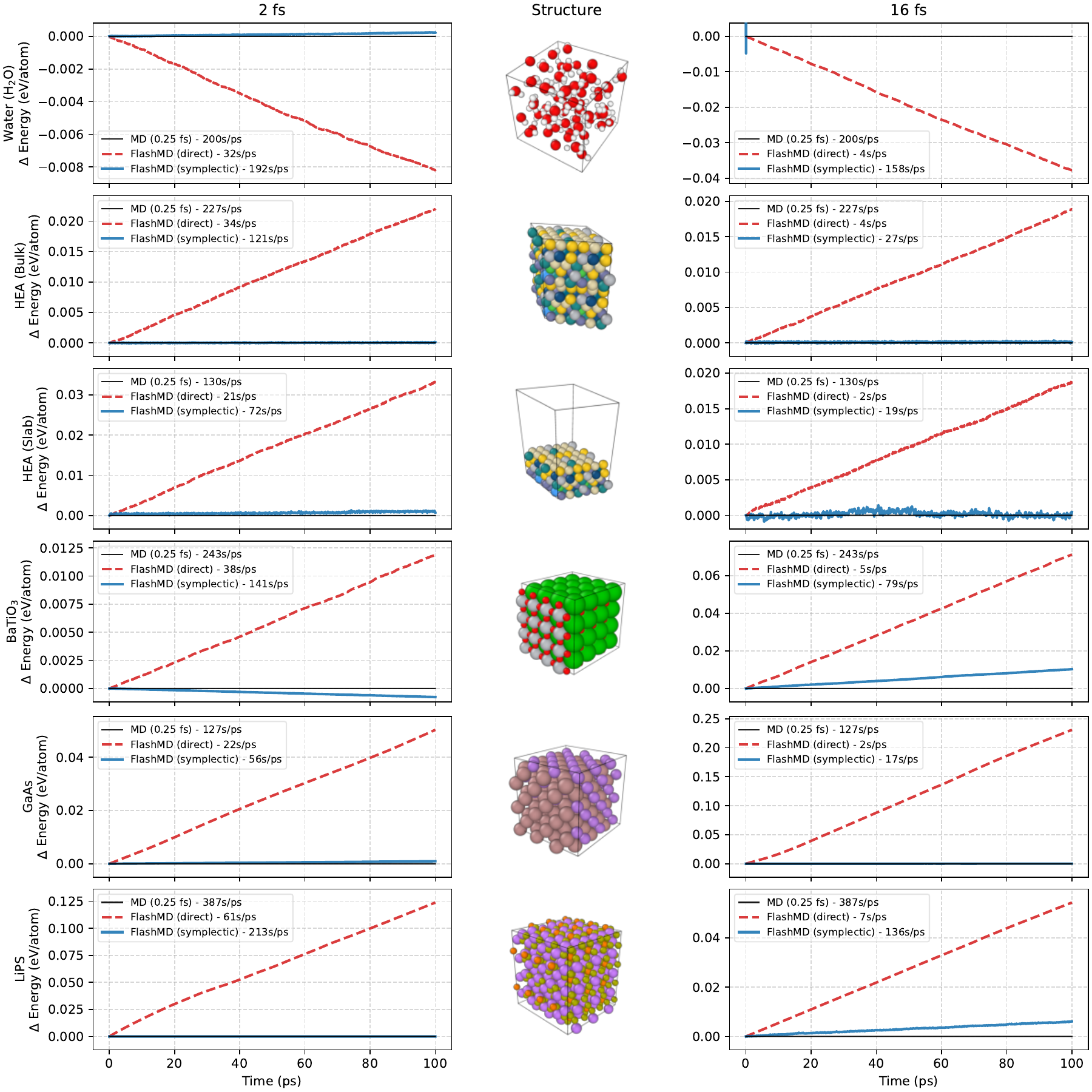}
\caption{Energy conservation comparison across six material systems. The plots display the drift in conserved energy per atom ($\Delta E$) over 100~ps, normalized to zero after 20~ps of equilibration. Left: 2~fs time step, right: 16~fs time step. Symplectic FlashMD (blue solid) exhibits superior stability compared to standard FlashMD (red dashed), maintaining negligible drift comparable to the 0.25~fs baseline (black solid), particularly at the larger 16 fs time step except for liquid water.
The legend includes wall-clock timings for the different simulation approaches, measured on a single NVIDIA H100. 
Note that 0.25~fs is an overly conservative choice for the systems considered here; more realistic time steps for these systems would be 
0.5~fs (for \ce{H2O}); 
2~fs (for HEAs);
2~fs (for BTO);
4~fs (for \ce{GaAs});
1~fs (for \ce{LiPS}).
}
    \label{fig:universal}
\end{figure*}

\subsubsection{Training details}

We trained our symplectic integrator models using the \texttt{metatrain} \cite{metatrain} framework, adapting the FlashMD architecture to output a scalar generating function $S^3_\theta(\bar{q},\bar{p})$. While the original FlashMD outputs new phase-space coordinates directly, we modified the architecture to include a scalar head, enabling the computation of symplectic updates via automatic differentiation. Specifically, the model predicts the changes in positions and momenta as gradients of the learned generating function as defined in eq. \ref{eq:s3}.

To enable stable training with double backward automatic differentiation, which is required to compute the updates from the scalar output, we carefully validated the architecture's numerical stability using 32-bit single-precision floating-point arithmetic. The shift from 64-bit precision significantly reduced memory overhead and computational cost without compromising model accuracy.

Furthermore, using single-precision floats enabled training on the same large datasets that FlashMD was trained on. This allows a fair comparison between the two models and enables the choice of a pre-trained FlashMD as an initial guess for the fixed-point iteration.

The models were trained by directly minimizing the Mean Absolute Error (MAE) between the predicted update steps and the ground truth changes derived from the reference dataset: $$\mathcal{L}(\theta)=\lambda_q\left\|\Delta\bq-\frac{\partial S^3_\theta(\bar{\bq},\bar{\bp})}{\partial\bar{\bp}}\right\|_1+\lambda_p\left\|\Delta\bp+\frac{\partial S^3_\theta(\bar{\bq},\bar{\bp})}{\partial\bar{\bq}}\right\|_1$$ where $\Delta\bp,~\Delta\bq,~\bar{\bp},$ and $\bar{\bq}$ are computed from the training trajectories. We weighted the position loss term by ${\lambda_q=50.0}$ relative to the momentum term $\lambda_p=1.0$ to account for scale differences. We found that the $L_1$ loss (MAE) outperformed $L_2$ (MSE) for our objective.

Training was conducted on 16 NVIDIA H100 GPUs using a distributed data-parallel strategy. We used a batch size of 16 per device and trained for 50 epochs. Optimization was performed using the Adam optimizer~\cite{kingma2014adam} with a learning rate of $1\times 10^{-4}$ and a linear warmup phase of 5\% of the total steps. The architecture has a node dimension of 1024 and edge features of dimension 256. We used three GNN layers.

\subsubsection{Simulation details}

All simulations were conducted in the NVT ensemble at 300~K. Temperature control was maintained using a stochastic velocity rescaling thermostat~\cite{buss+07jcp} with a time constant of $\tau=100~\text{fs}$. The total energy is not conserved in an NVT simulation but we can compute a conserved quantity by monitoring the heat balance with the thermostat~\cite{buss-parr07pre}.
We conducted simulations using both direct and symplectic FlashMD at 2~fs and 16~fs, and compare them with the reference MD trajectories, that were computed with an extremely small time step of 0.25~fs to ensure the highest possible level of accuracy.
Realistic time steps that could be used in MD simulations depend on the system.  

In each integration step, we solve eq. \ref{eq:s3} numerically via fixed-point iteration. We reformulate the problem as an update rule ${\bxbar_{n+1}=f(\bxbar_n)},$ with the state vector defined as $\bxbar_n=\begin{bmatrix}\bqbar_n&\bpbar_n\end{bmatrix}.$ The step function $f$ is defined by the relation $\bqbar_{n+1}=\bq+\frac{1}{2}\frac{\partial S^3_\theta}{\partial\bpbar_n}, \quad \bpbar_{n+1}=\bp-\frac{1}{2}\frac{\partial S^3_\theta}{\partial\bqbar_n}$, where $\bp$ and $\bq$ are the known coordinates at the current simulation step. At each step, we use FlashMD's prediction $\bx_0^\prime$ as an initial guess for the first iterate, from which we compute the initial midpoint $\bxbar_0.$ We repeat the fixed-point iteration until the difference between successive iterations falls below a predefined threshold: $\|\bxbar_{n+1}-\bxbar_n\|_2<\varepsilon.$ For our experiments, we determined that a threshold of $\varepsilon=1\times 10^{-4}$ provides sufficient accuracy.

The entire simulation workflow was implemented by coupling the i-PI universal force engine with a symplectic solver implemented in the FlashMD library.

For the simulations, we verified that the conserved energy remains as close to the velocity Verlet baseline as possible. In fig. \ref{fig:universal}, we compare the conserved energies of a velocity Verlet baseline, FlashMD, and our symplectic version over a trajectory of 100~ps. Except for water at 16~fs, the simulations remain close to the symplectic baseline with a simulation time of 100~ps.

\subsubsection{A pathological case: liquid water at 16 fs time steps}

It can be seen from Fig. \ref{fig:universal} that the symplectic universal model trained to predict 16 fs time steps fails in predicting the dynamics of liquid water. This failure can most likely be rationalized in terms of caustics of the two-point action $S(\bq,\bq')$.
An S-type generating function defines the map via
\begin{equation}
\bp = -\frac{\partial S}{\partial \bq}, \qquad \bp' = \frac{\partial S}{\partial \bq'}.
\end{equation}
Locally, this requires the endpoint map $(\bp,\bq) \rightarrow (\bq,\bq')$ to be well-defined, which in turn relies on the
invertibility of the Jacobian of $\bp$ as a function of $\bq'$, and as a result, the invertibility of the mixed Hessian
\begin{equation}
\frac{\partial^2 S}{\partial \bq \, \partial \bq'^{\top}}.
\end{equation}
At a caustic, this matrix becomes singular, the mapping $(\bp,\bq) \rightarrow (\bq,\bq')$ loses local invertibility, and distinct
classical trajectories correspond to the same pair $(\bq,\bq')$.
As a consequence, $S(\bq,\bq')$ becomes multi-valued (it features multiple branches) in the neighborhood of the caustic.

In this type of regime, any single-valued parametrization of the action must
implicitly select or blend branches, leading to inconsistent derivatives and ill-conditioned implicit updates.
While using $S^3(\bpbar, \bqbar)$ delays the first caustic relative to $S(\bq, \bq')$ (as discussed previously when motivating the choice of $S^3$), it does not remove it; for liquid water at $h=16$ fs we
likely probe a multi-branch region, consistent with the observed instability. Sampling the vicinity of a caustic might also be the reason for the moderate energy drifts observed in the simulations of lithium thiophosphate and barium titanate.
}

\rev{
\subsection{Accelerating the fixed-point iterations for dynamics propagation}

To efficiently solve the fixed-point equations arising from the implicit symplectic integration scheme, we explored Anderson~\cite{anderson1965iterative} acceleration. Unlike standard fixed-point iteration, which computes the next update based solely on the current state, Anderson acceleration utilizes a history of the previous $m$ iterates. At each step, the method computes a linear combination of these stored residuals to minimize the Euclidean norm of the error within the portion of phase space spanned by the history. This significantly reduces the number of iterations required to reach a specified tolerance. The efficiency of the solver is governed by the history depth ($m$) and the mixing parameter ($\beta$), which modulates the step size of the update.

\begin{figure}
    \centering
    \includegraphics[width=0.9\linewidth]{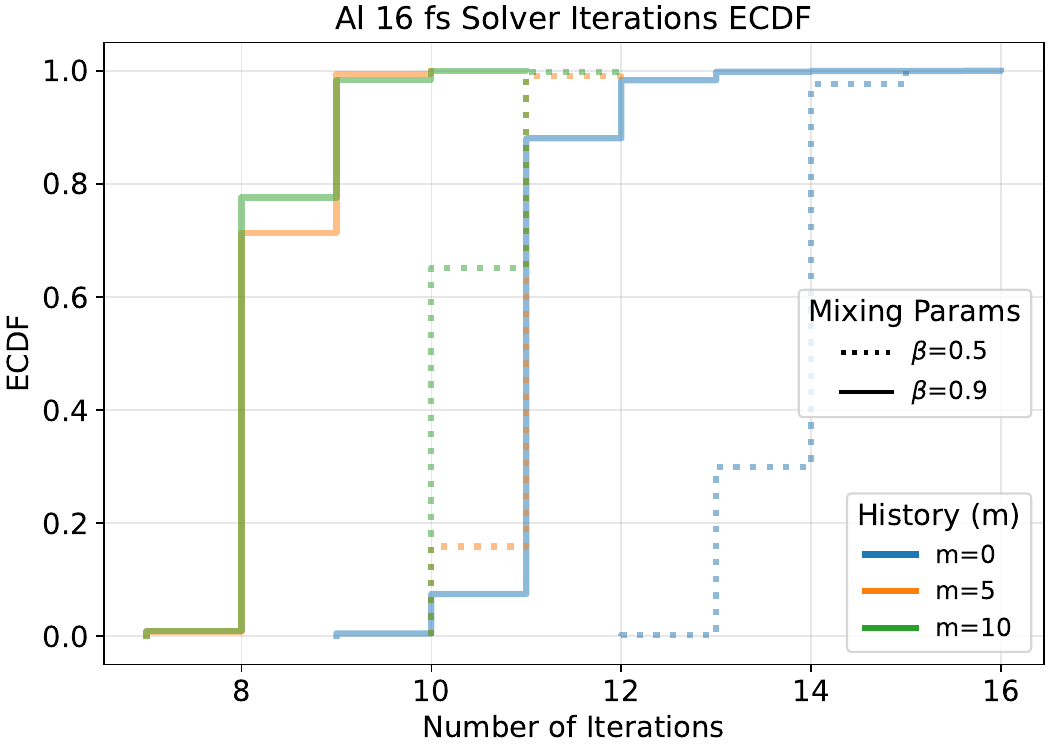}
    \caption{Empirical cumulative distribution function (ECDF) of solver iterations required for convergence using Anderson acceleration. Step functions show the observed iteration counts for varying history sizes ($m$) and mixing parameters ($\beta$).}
    \label{fig:anderson}
\end{figure}

We experimented with different history sizes. As illustrated in Fig. \ref{fig:anderson}, increasing the history size  generally improves convergence stability and reduces the computational cost per time step. During our experiments, we found that a history size of $m=5$ leads to fast convergence of the fixed-point solver. Increasing $m$ further to 10 was counterproductive for predicting 2~fs time steps but helped convergence of large system like \ce{BaTiO3} at 16~fs. 

\section{Summary of useful results from geometric numerical integration}

\subsubsection{Correspondence between sympletic maps and generating functions}

We will report results relative to the generating functions $S$ and $S^3$ (in the notation of Ref.~\cite{hairer2010geometric}), as these are the most directly relevant for our investigation. The former is also known as Hamilton's principal function, two-point action, Hamilton-Jacobi action; the second is also known as Poincaré generating function or midpoint generating function.

(Chapter 6 of Ref.~\cite{hairer2010geometric}, Theorem 5.1) A mapping $(\bp, \bq) \rightarrow (\bp', \bq')$ is symplectic if and only if there exists locally a function $S(\bq, \bq')$ such that
\begin{equation}
    \bp'^\top d\bq' - \bp^\top d\bq = dS.
\end{equation}
It follows immediately that
\begin{equation}
    \bp = - \frac{\partial S}{\partial \bq}, \quad \bp' = \frac{\partial S}{\partial \bq'}.
\end{equation}

(Chapter 6 of Ref.~\cite{hairer2010geometric}, Lemma 5.2) Similarly, if $S^3 = \bpbar \, \Delta \bq - S$, one has
\begin{equation}
    \Delta\bq^\top \bpbar - \Delta\bp^\top \bqbar = dS^3.
\end{equation}
and it follows that
\begin{equation}
    \Delta \bq = \frac{\partial S^3}{\partial \bpbar}, \quad \Delta \bp = - \frac{\partial S^3}{\partial \bqbar}.
\end{equation}

\subsubsection{Hamilton-Jacobi equation}

(Chapter 6 of Ref.~\cite{hairer2010geometric}, Theorem 5.5) If $S(\bq, \bq', t)$ is a smooth solution of
\begin{equation}
    \frac{\partial S}{\partial t} + H \Big( \frac{\partial S}{\partial \bq'}, \bq' \Big) = 0,
\end{equation}
and if the matrix $(\partial^2 S) / (\partial \bq \partial \bq'^\top)$ is invertible, there is a map defined by $(\bp, \bq) \rightarrow (\bp', \bq')$, where $\bp = - \partial S / \partial \bq$ and $\bp' = \partial S / \partial \bq'$, which is the flow of the Hamiltonian system with Hamiltonian $H(\bp, \bq)$.

For completeness, we also report the version of the Hamilton-Jacobi equation for $S^3(\bpbar, \bqbar, t)$ (Chapter 6 of Ref.~\cite{hairer2010geometric}, Theorem 5.7):
\begin{equation}
    \frac{\partial S^3}{\partial t} = H \Big( \bpbar - \frac{1}{2} \frac{\partial S^3}{\partial \bqbar}, \bqbar + \frac{1}{2} \frac{\partial S^3}{\partial \bpbar}  \Big).
\end{equation}

\subsubsection{Modified Hamiltonian and long-time behavior of symplectic integrators}

(Chapter 9 of Ref.~\cite{hairer2010geometric}, Theorem 8.1, restated and adapted to the current context) Let $\bp_t, \bq_t$ the momenta and positions given by the time-evolution of $\bp_0, \bq_0$ under a symplectic integrator. Then the modified Hamiltonian satisfies $\Tilde{H}(\bp_t, \bq_t) = \Tilde{H}(\bp_0, \bq_0) + \mathcal{O}(e^{-h_0/(2h)})$, where $h$ is the integration step and $h_0$ is a constant, and the real Hamiltonian satisfies $H(\bp_t, \bq_t) = H(\bp_0, \bq_0) + \mathcal{O}(h^p)$, where $p$ is the order of the symplectic method, over exponentially long times in $1/h$.

We note that, in our context, and as discussed under ``Monitoring molecular dynamics quality with symplectic predictors'', symplectic evolution and approximate energy conservation guarantee near-quantitative sampling of the relevant thermodynamic ensemble (within ergodicity assumptions). 
}

\end{document}